\documentclass[superscriptaddress,prx,twocolumn]{revtex4}
\usepackage{amsmath}
\usepackage{amssymb}
\usepackage{amscd}
\usepackage{graphicx}
\usepackage{bbm}
\usepackage{dsfont}
\usepackage{color}
\usepackage{datetime}
\usepackage{doi}

\longdate

\begin{document}
\title{Entanglements and Whitehead Products: Generalizing Kleman's Construction to Higher-Dimensional Defects}
\author{Gareth P. Alexander}
\email{G.P.Alexander@warwick.ac.uk}
\affiliation{Department of Physics, Gibbet Hill Road, University of Warwick, Coventry, CV4 7AL, United Kingdom.}
\affiliation{Centre for Complexity Science, Zeeman Building, University of Warwick, Coventry, CV4 7AL, United Kingdom.}
\author{Randall D. Kamien}
\email{kamien@upenn.edu}
\affiliation{Department of Physics and Astronomy, 209 South 33rd Street, Philadelphia, Pennsylvania 19104, USA.}
\affiliation{Department of Mathematics, 209 South 33rd Street, Philadelphia, Pennsylvania 19104, USA.}

\date{\today}

\begin{abstract}
We review the interpretation of Whitehead products in homotopy theory as an entanglement of topological defects in ordered media. 
\end{abstract}
\maketitle

The advantage of abstraction in mathematics and physics is that, in principle, we exchange remembering less with learning more: by showing that a whole class of phenomena can be interpreted as ``one big thing'' we can use one example as the explanation for the ``many things."   In the 1970s Maurice Kleman was instrumental in  introducing the use of homotopy theory in the characterization of defects across ordered media, synthesizing a myriad of observations ranging from flux lines in superconductors to Dirac monopoles to textures as one big idea.  Many of these ideas were captured in
 the 1976 paper with Toulouse~\cite{toulouse76}, the 1977 follow up~\cite{kleman77}, and the extension to translationally ordered media in 1978~\cite{kleman1978,kleman1978b}. Also in 1977 he described the topological implications of linked defects in terms of Whitehead products~\cite{kleman77b}, independently of closely similar results described for defect crossing by Po\'enaru \& Toulouse~\cite{poenaru77,poenaru79}.   
 
As a memorial to Maurice, we offer a small extension to this program by studying how defects interact topologically, as exemplified in the theory of linked defects in biaxial nematics~\cite{kleman77b,monastyrsky86}.  In particular, we will show how the inability to cross generic defects in the biaxial nematic arises from the same mathematical structure that explains how moving a hedgehog around a disclination line alters its apparent charge.  We will then use the same tool to construct the Hopf fibration out of two hedgehog-like textures or, more precisely, out of skyrmions.  As we have found innumerable times it is easy to fool ourselves into believing we have understood something new. We note that no particular result here is new but, rather, the result is that known results are related (unless we are mistaken).  We are at the same time inspired and humbled by Maurice's profound insights and hope that he would have appreciated this note.

The topological classification of defects in ordered media is based around the homotopy groups of the ground state manifold (GSM) \cite{rogula76,toulouse76,volovik76,kleman77,kleman1978,mermin79}, the space of equivalent but distinct ground states of the ordered phase. One of its key results is that this classification mirrors Landau's symmetry breaking approach to phase transitions. If the symmetry breaking is from the high temperature group $G$ to the low temperature group $H$ then the ground state manifold is the coset space $G/H$ and the defects are classified by conjugacy classes of the homotopy groups $\pi_{n}(G/H)$. 
To be more precise, the homotopy groups give a local characterization of defects in the ordered phase, based upon measurements of the texture on a Burgers circuit surrounding the defect. For line-like defects, such a circuit is a loop, or circle ($S^1$), surrounding the line and the measurement corresponds to a map $S^1 \to G/H$. Any continuous deformation of the texture describes a homotopy of this map and the defect is associated to its equivalence class. The set of distinct defects is therefore given by the fundamental group $\pi_1(G/H)$.  Among the mathematical details, which play a role in this note, is that homotopy groups are defined with respect to a base point.  The loops represented by the elements of $\pi_1(G/H)$ must all begin and end at the same, fixed point.  The higher homotopy groups $\pi_n(G/H)$ classify maps from the $n$-dimensional sphere ($S^n$) to the GSM.  For instance, when $\pi_2(G/H)$ is not the identity group there must be point-like defects in a three-dimensional sample: the hedgehog in the Heisenberg model or in nematic liquid crystals.  Nematics are especially rich, topologically.  We have $G=SO(3),\,H=D_{\infty}$ and find that the ground state manifold is $G/H\cong\mathbb{RP}^2$, the real projective plane. The homotopy groups are $\pi_1(\mathbb{RP}^2) = \mathbb{Z}_2$, $\pi_2(\mathbb{RP}^2) = \mathbb{Z}$ and $\pi_3(\mathbb{RP}^2) = \mathbb{Z}$.

\begin{figure*}[t]
    \centering
    \includegraphics[width=\textwidth]{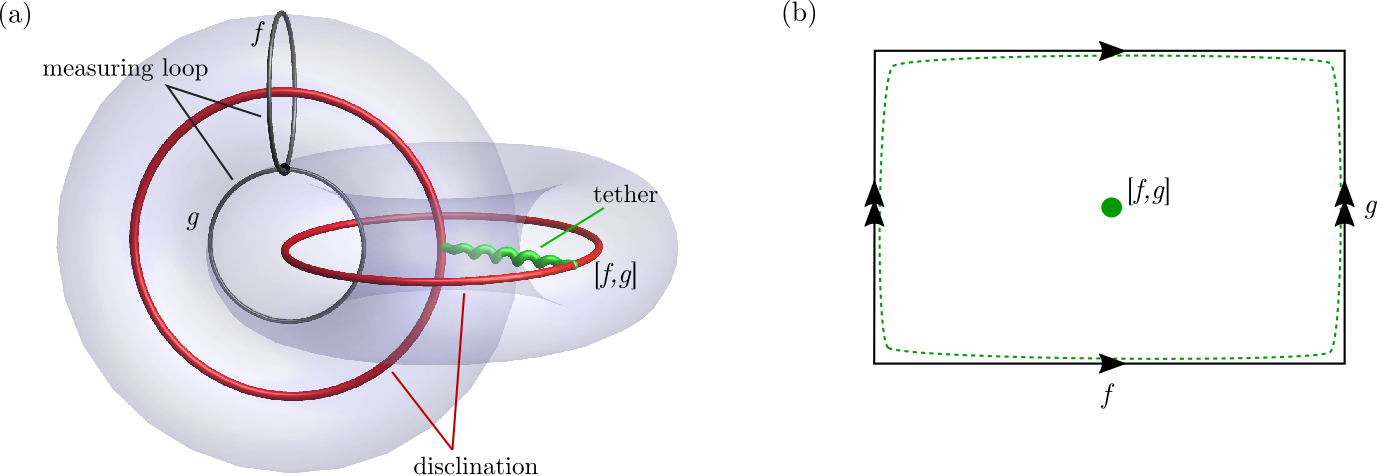}
    \caption{(a) Linked defect loops (red) are in general connected by a third defect (green). Each defect can be encircled by a measuring loop, recording textures $f,g: S^1 \to G/H$. The connecting tether is characterised by the homotopy class of their commutator $[f,g] = fgf^{-1}g^{-1}$, an example of the Whitehead product. (b) If the torus surrounding either defect loop is cut open along the two measuring loops it can be laid flat as a rectangle with boundary the two cut measuring loops. A loop following around the boundary of the rectangle (dashed green) records the commutator $[f,g]$, which is also the homotopy class of the connecting defect tether.}
    \label{fig:linked_loops}
\end{figure*}

The homotopy groups provide a means of labelling individual defects by topological invariants and, through their group structure, a framework for understanding, or indeed predicting, how multiple defects combine, or single defects break apart. There are also interactions between defects of different types and it is important to identify their topological character too. A noteable example of this arises in liquid crystal colloids where the interplay is between line and point defects~\cite{terentjev95,poulin97,musevic06,ravnik09,lapointe09,lintuvuori10,tkalec11}. Interactions between defects of different types reflect, at the level of topology, relations between the homotopy groups that were first described by J.~H.~C.~Whitehead~\cite{jhcwhitehead41} and which underpin the crossing phenomenon described by Po\'enaru and Toulouse \cite{poenaru77,poenaru79}. These were also described by Kleman for the specific setting of linked defect loops~\cite{kleman77b}. 

As shown in Fig.~\ref{fig:linked_loops}(a), we can consider two linked defect loops.  To isolate the singular defects, we sheath each loop with a measuring torus and consider maps from the torus to the GSM.  The torus has two cycles; any meridional cycle encircles the disclination loop inside the torus and a measurement along it detects the homotopy class in $\pi_1(G/H)$ corresponding to that defect. Similarly, the longitudinal cycle encircles the other disclination loop (outside the torus) and records its homotopy class. Now suppose, as in Fig.~\ref{fig:linked_loops}(b), that the torus is cut open along these cycles and laid flat as a rectangle in the plane whose boundary is made up from the two cycles we have cut. This boundary of the rectangle is another closed loop and a measurement of the texture along it furnishes another homotopy class in $\pi_1(G/H)$. This measurement is readily seen to be a `commutator' of the measurements made on the meridional and longitudinal cycles. That is, if the textures on the meridional and longitudinal cycles yield measurements $f,g:S^1\rightarrow G/H$, respectively, then the boundary of the rectangle produces the measurement $fgf^{-1}g^{-1}$, which has the structure of a commutator. Only if the homotopy class of this induced map on the boundary of the rectangle is trivial can it be extended to a continuous map on the entire rectangle. If it is not trivial then the torus is `pierced' by a defect with this homotopy type. As this must be true of any torus enclosing the disclination loop, we see that this new defect forms a tether connecting the two loops.  Note that if $\pi_1(G/H)$ is abelian then the commutator is always the identity and no defect is required.  In the famous case of biaxial nematics where $G/H\cong \mathbb{Q}_8$ (the unit quaternions) $\pi_1(G/H)$ is non-abelian and tethers form \cite{toulouse1977,kleman77b,mermin79}. Topological defects become topologically entangled!

The homotopy class of the connecting tether is known as the Whitehead product of the homotopy classes of the two disclination loops it connects \cite{kleman77b,poenaru77}. Although the definition and properties of Whitehead products are available in standard textbooks~\cite{Whitehead,Hatcher}, we provide a brief summary to keep our discussion self-contained. The Whitehead product is a map between homotopy groups, $\pi_1(G/H)\times\pi_1(G/H)\rightarrow \pi_1(G/H)$.  We connect two circles so that they have the same base point and denote this by $S^1 \vee S^1$. Any two based maps $f,g$ define a map $S^1 \vee S^1 \to G/H$ by identifying the base point on the two $S^1$s. By gluing a disc $D^2$ to the scaffold $S^1 \vee S^1$ we can create a torus $S^1 \times S^1$. This attaching map induces a map $[f,g]: \partial D^2 \to G/H$, whose homotopy class is the Whitehead product of the classes $\alpha,\beta\in\pi_1(G/H)$ represented by the maps $f,g$. 

This construction admits a natural generalisation to the case where the starting measuring circuits are a $k$-sphere and a $l$-sphere. Gluing the boundary of a $(k\!+\!\ell)$-disc to the scaffold (technically known as ``the wedge'') $S^k\vee S^\ell$ so as to produce a torus $S^k\times S^\ell$ defines a map $[f,g]:\partial D^{k+\ell}\cong S^{k+\ell-1}\rightarrow G/H$ whose homotopy type is the Whitehead product of the classes $\alpha,\beta$ represented by the initial measurements $f,g$. Thus the Whitehead product induces an algebra structure on the homotopy groups $\pi_*(G/H)$, which manifests itself in the topological character of interactions between defects of different types \cite{poenaru79}.   In short, there are maps $$[f,g]:\pi_k(G/H)\times\pi_\ell(G/H) \rightarrow \pi_{k+\ell-1}(G/H).$$  When the result is a non-trivial element of $\pi_{k+\ell-1}(G/H)$ it implies that the map from the attached disk $D^{k+\ell}$ to $G/H$ must have a singularity -- a new defect!

\begin{figure*}[!t]
\centering
\includegraphics[width=\textwidth]{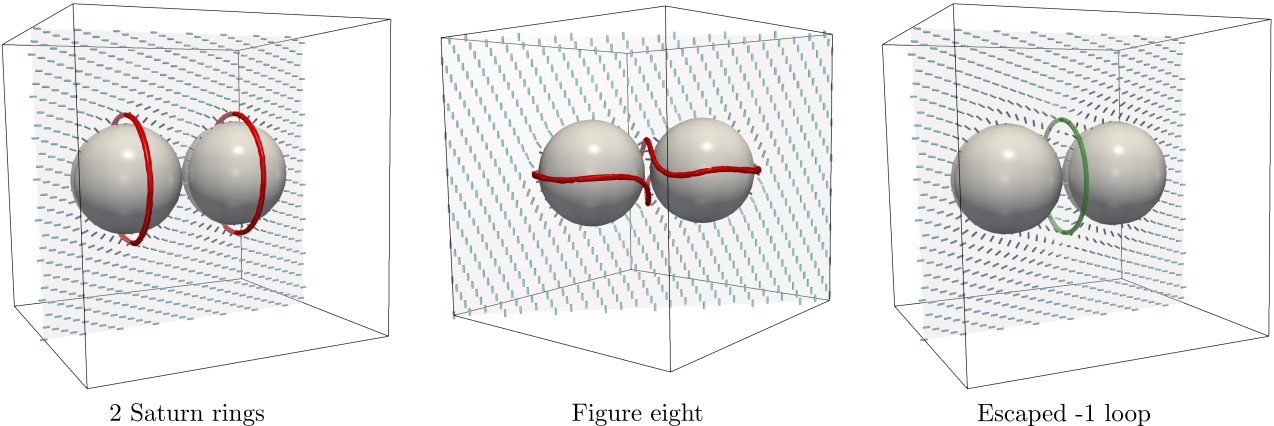}
\caption{Colloidal inclusions illustrate the topological interaction between line defects ($\pi_1$) and point defects ($\pi_2$). Normal anchoring boundary conditions make the colloids act like point defects to the surrounding liquid crystal, which can compensate for it in several different ways; with point defects, with separate disclination loops, with a single disclination loop, or with a defect free texture. The process that converts the single defect loop (figure eight structure) into the defect free state (escaped $-1$ loop) is an instance of the Whitehead product between $\pi_1(\mathbb{RP}^2)$ and $\pi_2(\mathbb{RP}^2)$.}
\label{fig:colloids}
\end{figure*}

For two disclination loops, both defects are of the same type and we found that the Whitehead product was only interesting when $\pi_1(G/H)$ was non-abelian.  The Whitehead product, however, can be used to mix and match defects of different type, for instance point defects and defect loops.  Even if $\pi_1(G/H)$ is abelian ($\pi_n(G/H)$ is always abelian for $n\ge 2$) we will see that the action of a line on a point can be interesting -- indeed it is the reason that when we carry a nematic hedgehog through a disclination loop its charge ``changes sign''~\cite{mermin79,alexander12}.  
The simplest example of this is the situation encountered in colloidal liquid crystals, where the colloids act as surrogate point defects. This purely topological interaction is at the heart of the topological classification of both disclination loops and point defects~\cite{janich87,nakanishi88,bechluft-sachs99,alexander12,copar11,copar11b,copar12,machon2016}; see Fig.~\ref{fig:colloids}.  

Colloidal particles are usually prepared so that they present homeotropic boundary conditions to the liquid crystal host and as a result they behave like unit charge point defects. This charge may be compensated in two characteristic ways; either by an accompanying point defect in the bulk liquid crystal to give a `hedgehog dipole' configuration \cite{poulin97}, or by a disclination loop surrounding the particle in a quadrupolar `Saturn ring' configuration \cite{terentjev95}. The Saturn ring configuration demonstrates that disclination loops may be associated with a homotopy class in $\pi_2(\mathbb{RP}^2)$, corresponding to the fact that they are able to compensate the charge of the colloid, in addition to a class in $\pi_1(\mathbb{RP}^2)$, corresponding to any measurement made along a loop encircling the defect line. However, this element of $\pi_2(\mathbb{RP}^2)$ is {\sl not} a homotopy invariant of the disclination loop~\cite{janich87,nakanishi88,bechluft-sachs99,copar11b,alexander12}. 

This feature is reflected in the structures that form when two Saturn ring colloids form entangled structures, for it is possible to produce configurations where the two colloids are surrounded by a single disclination loop~\cite{ravnik07} or by no topological defect in the liquid crystal at all \cite{tkalec09}, as shown in Fig.~\ref{fig:colloids}. In the former case the colloids are interpreted as having the same charge ($\pm 1$) and the disclination loop as having hedgehog charge $\mp 2$ to compensate. In the latter case, the absence of any defects in the liquid crystal implies that the colloids must be identified with opposite hedgehog charges, one positive and one negative. In the topological classification of disclination loops, this dichotomy reflects the fact that the hedgehog charge of a disclination loop is only a homotopy invariant $\text{mod}\;2$ \cite{janich87,copar11b,alexander12}. The process that converts one two-colloid texture into the other and gives a representation of the homotopy changing the hedgehog charge of a disclination loop by 2, is a realisation of the Whitehead product between $\pi_1(\mathbb{RP}^2)$ and $\pi_2(\mathbb{RP}^2)$. 

\begin{figure}[t]
\centering
\includegraphics[width=75mm]{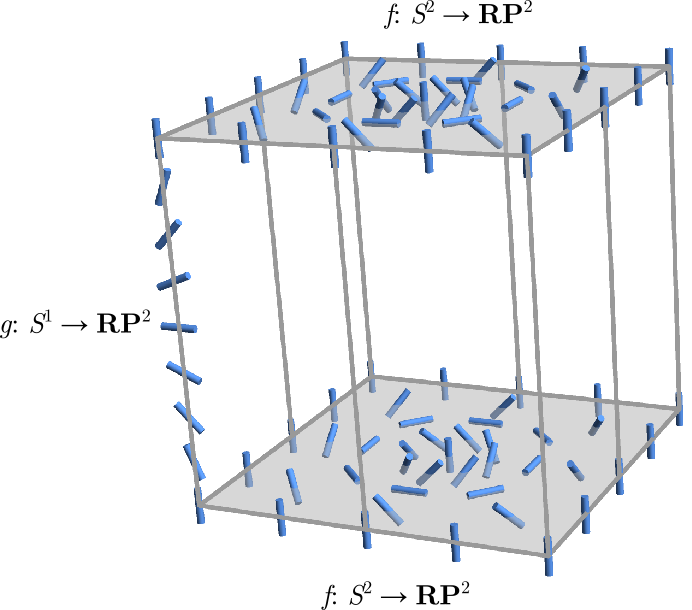}
\caption{Construction of the Whitehead product between $\pi_1$ and $\pi_2$. A texture $f: S^2 \to \mathbb{RP}^2$ representing a homotopy class in $\pi_2$ covers the bottom and top faces of a cube, while the sides are filled with a texture $g: S^1 \to \mathbb{RP}^2$. Taken together the surface of the cube is a map $[f,g]: S^2 \to \mathbb{RP}^2$ whose homotopy class is the Whitehead product of the two separate textures.}
\label{fig:whitehead_pi1_pi2}
\end{figure}

To see this, we need a little help with the visualisation, Fig.~\ref{fig:whitehead_pi1_pi2}.  To start, instead of drawing the two-sphere as embedded in three dimensions, we instead draw a square in the plane and identify all the points on all the edges.  Then maps from $S^2$ to any GSM are maps from the square that map the entire boundary to one point.  To build the Whitehead product of $\pi_2(G/H)$ and $\pi_1(G/H)$, we represent the circle as the unit interval, $[0,1]$ and extend the square into the third dimension along the unit interval.  The top of the open box is now covered with the same square representation of the $S^2$.  We now have a two-dimensional surface surrounding the solid cube, homotopic to $S^2$ and $D^3$, respectively.  On the top and bottom we have a map $f:S^2\rightarrow G/H$ with class $\alpha\in\pi_2(G/H)$ and along the sides (which are all one point in the $xy$-plane!) we have a map $g: S^1\rightarrow G/H$ with class $\beta\in\pi_1(G/H)$.  
When $G/H=\mathbb{RP}^2$, this becomes interesting.  In order to measure the charge $\alpha$ of a point defect we need to lift the nematic director field to a unit vector field creating a map from $S^2\rightarrow S^2$ from which we can calculate its degree \cite{alexander12}.  There are two choices when choosing the vector field (``up'' or ``down'') but, when mapping from $S^2$ the choice can always be made consistently.  The complication in a nematic is that to compare two point defects at {\sl different} locations one must also consistently lift the director field along a tubular region {\sl connecting} the two points.  If there is a line defect lurking then the relative signs of the charges depend on the path taken to make the comparison.  This is where the map $g$ comes in.  As we move from the bottom to the top of the cube, the map $g$ can rotate the vector field by an odd multiple of $\pi$ or an even multiple of $\pi$.  The former implies that we have circled a disclination loop while the latter indicates that no line defects spoil the comparison process.  For instance, a single hedgehog in a nematic can lift to either an outward pointing hedgehog with charge $+1$ or an inward pointing hedgehog with charge $-1$.  Since the map $f$ looks the same for both the top and bottom surfaces of the cube can have the same map $f$ even if they lift to opposite charges.  The sides of the cube determine the relative lift of the top and bottom.  If there are no line defects then the top and bottom have the same charge; if there are line defects they have opposite charge.  Where does this get us?  Considering the boundary of the cube and the texture that we have built onto it with $f$ and $g$, we have generated a {\sl new} map from $[f,g]:S^2\rightarrow \mathbb{RP}^2$.   When we calculate the degree we get three contributions: the degree of the top surface map, the degree of the bottom surface map, and the degree of the map induced on the sides.  The last one does not contribute because the director field moves only along a one-dimensional curve on $S^2$.  We thus measure the charge of the top plus the charge of the bottom (both with respect to an outward pointing surface normal) --  if there is no twist by $\pi$ on the sides then the two charges cancel; if there is a twist defect they add.  Were we to start with a single hedgehog, the presence of the line defect would imply that $[f,g]$ has degree $\pm 2$.  If we were to try to fill the cube we would find that there must be a singularity somewhere inside with charge $\pm 2$. This corresponds directly to what we described earlier for the Whitehead product of two elements of $\pi_1$, where we found a defect -- the connecting tether -- in the interior of the square (Fig.~\ref{fig:linked_loops}). The Whitehead product there arose from the attaching map of the square to a wedge of two circles so as to create a torus $S^1\times S^1$. In the present case, the corresponding construction attaches the cube to a wedge $S^1\vee S^2$ to create the three-manifold $S^1\times S^2$ and here lies the difference: whereas we could realise $S^1\times S^1$ directly as embedded in $\mathbb{R}^3$ (surrounding a pair of linked defects), $S^1\times S^2$ does not embed in $\mathbb{R}^3$. The exact analogous construction of linked defects connected by a tether exists only in $\mathbb{R}^4$ (and higher) so that in ordinary three-dimensional space we catch only a glimpse of the full construction. 

\begin{figure}[tb]
\centering
\includegraphics[width=80mm]{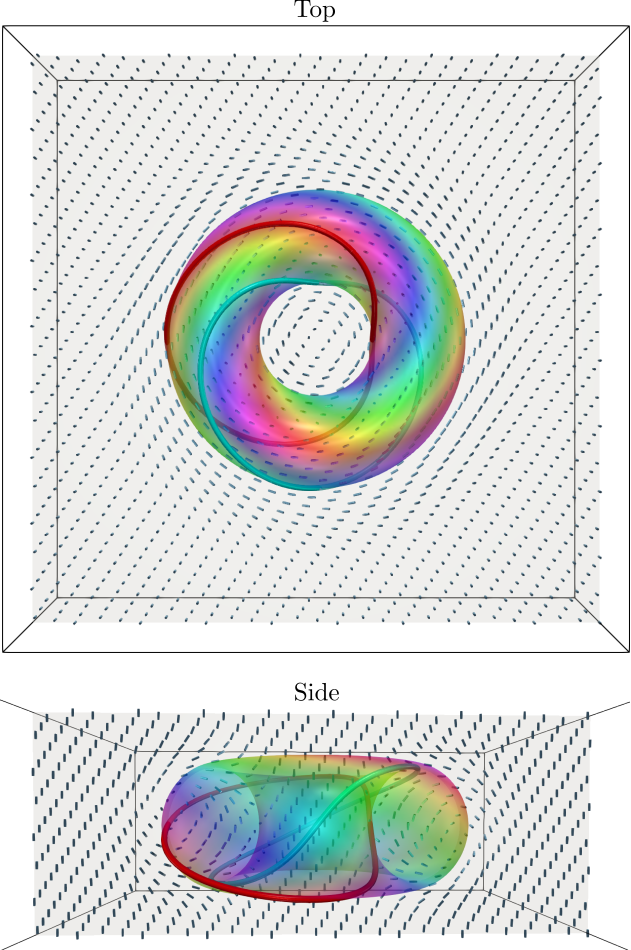}
\caption{The Hopf texture is a nonsingular director configuration associated with the homotopy group $\pi_3(\mathbb{RP}^2) = \mathbb{Z}$. It is characterised by the linking of preimages of any two director orientations. In the figure, we show the preimages of ${\bf n}_1 = {\bf e}_x$ (red) and ${\bf n}_2 = {\bf e}_y$ (cyan) as thickened tubes; they have linking number $+1$, which is the Hopf invariant. The colored surface is the preimage of all horizontal directions, ${\bf n}_i \cdot {\bf e}_z = 0$, known as a Pontryagin-Thom surface~\cite{chen13}.}
\label{fig:hopf}
\end{figure}

Is there more?  One might think that once we have explored point and line defects in the nematic, there would be nothing left.  But nematic liquid crystals also enjoy {\sl textures} -- nonsingular yet topologically nontrivial configurations.  The Hopf texture is the canonical and most famous example of this~\cite{chen13}. If we consider two different directions on the nematic GSM, then we can ask,``where in the sample does the director point along ${\bf n}_1$ and ${\bf n}_2$?'' First note that if the nematic were uniform in space then it only points in one direction and, generically, the preimages of the ${\bf n}_i$ are empty!  This is the null topological structure.  However, as depicted in Fig.~\ref{fig:hopf} it is possible to have a more complex texture.  Since the GSM, $\mathbb{RP}^2$, is two-dimensional and our sample is three-dimensional the preimage will be one-dimensional and so, we expect interesting preimages to be a collection of closed loops and infinite lines (the latter can be turned into closed loops by adding the ``point at infinity'' making the sample into the compact three-sphere $S^3$).  When two distinct preimages of ${\bf n}_1$ and ${\bf n}_2$ are each single closed loops that are {\sl linked}, this is the Hopf fibration.  Note that two preimage loops can never cross under smooth deformations: if they did the two preimages would overlap at some point and then the director at that point would be undefined creating a singularity.  Thus, {\sl even without defects} the Hopf texture is topologically distinct from the null, uniform configuration.   This is, however, no different from the simple hedgehog: we characterize the topological charge of the hedgehog by considering the smooth configuration of the director field on a measuring sphere.  When that charge is non-vanishing, we conclude that there must be a singularity inside the two-sphere for if the filling of the sphere is also smooth, we can contract the measuring sphere to a point and get a vanishing charge.   In the case of the Hopf texture we are considering smooth director configurations on $S^3$ and measuring a topological charge.  Were we to imagine the nematic living in a four-dimensional sample then the filling of $S^3$ would be forced to have a singularity if the charge is non-vanishing.  In lower dimensions, the liquid crystal topology has an ``extrinsic'' character -- a surface measurement implies something about the bulk since the surface is embedded in the sample.  In the case of the Hopf fibration, there is no notion of an embedding -- the sample {\sl is} $S^3$, created by taking $\mathbb{R}^3$ and adding the point at infinity!

The Hopf texture can also be interpreted in terms of the Whitehead product.  If we embed $S^3$ in the four-dimensional space parameterized by $(w,x,y,z)\in\mathbb{R}^4$ then the equation for $S^3$ is 
\begin{equation}
1=w^2+x^2 + y^2 + z^2  = \zeta_1\bar\zeta_1 + \zeta_2\bar\zeta_2
\end{equation}
where we have grouped $w$ and $x$ ($y$ and $z$) into a complex number $\zeta_1=w + ix$ ($\zeta_2=y+iz$).  If we fix the magnitude of $\zeta_1$ to $r_0\in (0,1)$ then we have a general solution $\zeta_1=r_0 e^{i\theta_1}$ and $\zeta_2 = \sqrt{1-r_0^2} e^{i\theta_2}$.  We can thus divide the three sphere into two solid tori bounded by the common torus parameterized by the two angles $\theta_1$ and $\theta_2$.  The inner solid torus is when $\vert\zeta_1\vert \le r_0$ and the outer solid torus is the complement.   Each solid torus has a disc $D^2$ as a cross section and is equivalent to $S^1\times D^2$.  

\begin{figure}[b]
\centering
\includegraphics[width=80mm]{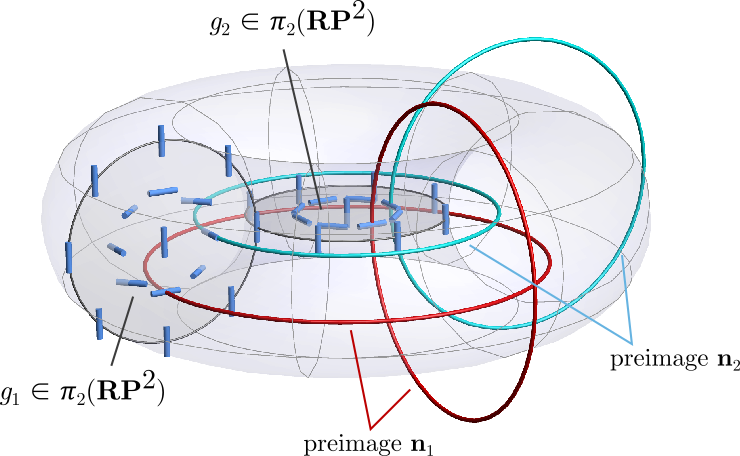}
\caption{Illustration of the Whitehead product of two elements of $\pi_2$. The three-sphere ($\mathbb{R}^3$ with the point at infinity added) can be split into a union of two solid tori, with common boundary a torus (light grey). Each cross-sectional disc of the `interior' solid torus is filled with a texture $g_1: S^2 \to \mathbb{RP}^2$ representing a homotopy class in $\pi_2$ and, similarly, each cross-sectional disc of the `exterior' solid torus is filled with a texture $g_2: S^2 \to \mathbb{RP}^2$. The whole texture represents a map $S^3 \to \mathbb{RP}^2$ whose homotopy class is the Whitehead product of the two elements of $\pi_2$. This homotopy class can be determined by calculating the linking of preimages of any two directions, as shown. In the example shown, the two elements of $\pi_2$ both have charge $\pm 1$ and the linking number of preimages is $\pm 2$.}
\label{fig:pi2product}
\end{figure}

To build a Hopf texture, choose an element $g_1\in\pi_2(S^2)$ -- usually associated with a hedgehog configuration.  As before, stretch the south pole out to a circle and identify all points of the circle with each other.  Thus we have a two-disc with the boundary identified representing our two-sphere.  Nematic configurations then correspond to ``baby Skyrmions'' on the disc and, without loss of generality, we can choose the director to point south on the boundary.  Now choose a second element $g_2\in\pi_2(S^2)$.  Fill the cross section of the inner solid torus with $g_1$ and the cross section of the outer solid torus with $g_2$.  By construction, on the common boundary torus the director field is uniformly pointing south.  We have now created a smooth texture on $S^3$ by using two elements of $\pi_2(S^2)$.  The Whitehead construction $\pi_2\times\pi_2\rightarrow \pi_3$ tells us that there is a way of filling the inside of $S^3$ with a texture: the two two-discs can each be morphed into the two squares $[0,1]\times[0,1]$ with the boundary indentified.  Wedging one square with the other gives us the boundary of the {\sl four-dimensional} box, $[0,1]^4$, equivalent to a four-dimensional ball with $S^3$ as its boundary.  Reading the topology of the boundary we have now created a smooth configuration on $S^3$ -- a map from $\mathbb{RP}^2$ to $S^3$ -- a texture!  What is its charge?  Look at the preimage of two different directions ${\bf n}_1$ and ${\bf n}_2$ (don't choose either pointing south!).  In the simplest case, let $g_1$ and $g_2$ each be the degree one element of $\pi_2(S^2)$, {\sl i.e.}, the simple hedgehog.  Since they are degree one, each has one signed occurence of ${\bf n}_1$ and ${\bf n}_2$ each of which has a preimage in both the inner and outer solid tori.  The preimages are all closed loops that wind through the solid tori and so we see that the ${\bf n}_1$ loop of the inner solid torus is linked with the ${\bf n}_2$ loop of the outer solid torus and {\sl vice versa}!  We have generated a texture on $S^3$ where two preimages are linked twice, giving Hopf charge $2$.  If we had chosen two elements of $\pi_2(S^2)$ with degree $p$ and $q$, then this construction creates a total linking of $2pq$.  We remark that this total linking and value for the Hopf invariant is the same as arises for the helicity of a solenoidal vector field~\cite{moffatt1969,moffatt1992} where essentially the same construction as in Fig.~\ref{fig:pi2product} is used.  It has been suggested that using a relative Whitehead product it would be possible to construct odd-integer Hopf charges \cite{machon}.

While we have not created any new defect structures in this note, we have shown that Maurice's original insight into the linking of disclination loops in the biaxial nematic is actually the basis for higher-dimensional considerations in nematics.  Whether we or someone else can extend these constructions for other liquid crystalline phases is an open question.

We gratefully acknowledge beneficial discussions with Dan Beller, Bryan Gin-ge Chen, Fred Cohen, Tom Machon, and Sabetta Matsumoto.

\end{document}